\documentclass[prd,aps,nofootinbib,twocolumn,linenumber,superscriptaddress,preprintnumbers,balancelastpage,longbibliography]{revtex4-1}
\usepackage{ae,aecompl}
\usepackage{graphicx}	
\usepackage{amsmath}	
\usepackage{graphicx}
\usepackage{dcolumn}
\usepackage{bm}
\usepackage{graphics}
\usepackage{afterpage}
\usepackage{float}
\usepackage{subfigure}
\usepackage{rotating}
\usepackage{multirow}
\usepackage{tabularx}
\usepackage{booktabs}
\usepackage{multirow}
\usepackage{fancyheadings}
\usepackage{fancyhdr}
\usepackage{soul}

\usepackage{theorem}
\usepackage{moreverb}
\usepackage{euscript}
\usepackage{psfrag}
\usepackage{slashed}
\usepackage{mathtools}
\usepackage{makecell}
\usepackage[flushleft]{threeparttable}
\usepackage{adjustbox}
 
\usepackage[utf8]{inputenc}
\usepackage[english]{babel}

\usepackage{dcolumn}
\usepackage{bm}
\usepackage[mathlines]{lineno}

\DeclareUnicodeCharacter{2212}{-}

\usepackage[colorlinks=true
,urlcolor=blue
,anchorcolor=blue
,citecolor=blue
,filecolor=blue
,linkcolor=black
,menucolor=blue
,linktocpage=true
,pdfproducer=medialab
,pdfa=true
]{hyperref}
\usepackage{listings}
\usepackage{color,xcolor}
\usepackage{placeins}
\usepackage{array}

\usepackage{academicons}

\newcommand{\orcidlink}[1]
{\begingroup
  \hypersetup{hidelinks}\href{https://orcid.org/#1}{\includegraphics[width=10pt]{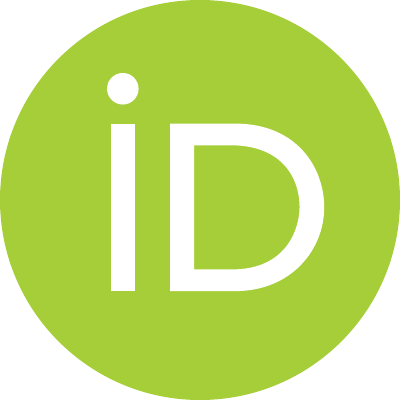}
} \endgroup}

\renewcommand{\arraystretch}{1.8}
\newcolumntype{C}[1]{>{\centering\let\newline\\\arraybackslash\hspace{0pt}}m{#1}}

\lstdefinestyle{python}{
  belowcaptionskip=1\baselineskip,
  breaklines=true,
  frame=L,
  xleftmargin=\parindent,
  language=Python,
  showstringspaces=false,
  basicstyle=\small\ttfamily,
  morekeywords={models, lambda, forms,True,False,None},
  keywordstyle=\bfseries\color{deepgreen!40!black},
  commentstyle=\itshape\color{gray},
  identifierstyle=\color{black},
  stringstyle=\color{deepred},
  rulecolor=\color{gray},
}

\newcommand{\DM}{{\scriptscriptstyle \text{DM}} }
\newcommand{\sigmav}{$\langle \sigma v \rangle$}

\begin{document}

\title{TeV gamma-ray sensitivity to velocity-dependent dark matter models \\ in the Galactic Center}

\author{Alessandro Montanari \orcidlink{0000-0002-3620-0173}}
\email{amontanari@lsw.uni-heidelberg.de}
\affiliation{\!\!\mbox{ \footnotesize Landessternwarte, Universität Heidelberg, Königstuhl, D 69117
Heidelberg, Germany}\,\,\,\vspace{0.7ex}}
\affiliation{\!\!\mbox{ \footnotesize IRFU, CEA, D\'epartement de Physique des Particules, Universit{\'{e}} Paris-Saclay, F-91191 Gif-sur-Yvette, France}\,\,\,\vspace{0.7ex}}
\author{Oscar Macias \orcidlink{0000-0001-8867-2693}} 
\email{o.a.maciasramirez@uva.nl}
\affiliation{\!\!\mbox{ \footnotesize Department of Physics and Astronomy, San Francisco State University, San Francisco, CA 94132, USA}\,\,\,\vspace{0.7ex}}
\affiliation{\!\!\mbox{ \footnotesize GRAPPA Institute, University of Amsterdam, 1098 XH Amsterdam, The Netherlands}\,\,\,\vspace{0.7ex}}
\author{Emmanuel Moulin \orcidlink{0000-0003-4007-0145}}
\email{emmanuel.moulin@cea.fr}
\affiliation{\!\!\mbox{ \footnotesize IRFU, CEA, D\'epartement de Physique des Particules, Universit{\'{e}} Paris-Saclay, F-91191 Gif-sur-Yvette, France}\,\,\,\vspace{0.7ex}}

\date{\today}
\newcommand{\aap}{Astronomy and Astrophysics}
\newcommand{\mnras}{Monthly Notices of the RAS}

\begin{abstract}
The center of the Milky Way is a prime site to search for signals of dark matter (DM) annihilation due to its proximity and expected high concentration of DM.
The amplification of the dispersion velocity of DM particles in the Galactic center (GC), caused by baryonic contraction and feedback, makes this particular region of the sky an even more promising target for exploring velocity-dependent DM models.
Here we demonstrate that current GC observations with the H.E.S.S. telescope, presently the most sensitive TeV-scale gamma-ray telescope in operation in this region of the sky, set the strongest constraints on velocity-dependent annihilating DM particles with masses above 200 GeV. For p-wave annihilations, they improve the current constraints by a factor of $\sim$4 for a DM mass of 1 TeV.
For the spatial distribution of DM, we use the results of the latest FIRE-2 zoom cosmological simulation of Milky Way-size halos. In addition, we utilize the newest version of the GALPROP cosmic-ray propagation framework to simulate the Galactic diffuse gamma-ray emission in the GC. 
We have found that p-wave (d-wave) DM particles with a mass of approximately 1.7 TeV and annihilating into the $W^+W^−$ channel exhibit a velocity-weighted annihilation cross-section upper limit of 4.6$\times$ 10$^{-22}$ cm$^3$s$^{-1}$ (9.2$\times$10$^{-17}$ cm$^3$s$^{- 1}$) at a 95\% confidence level. This is about 460 (2$\times$ 10$^{6}$) times greater than the thermal relic cross-section for p-wave (d-wave) DM models.

\end{abstract}

\pacs{PACS}
\maketitle

\section{Introduction}\label{sec:Introduction}
A substantial body of observational evidence in astrophysics and cosmology suggests that non-baryonic dark matter (DM) dominates the matter content of the 
universe~\cite{Zwicky:1933gu,1974Natur.250..309E,Rubin:1978kmz,Trimble:1987ee,Planck:2018vyg}. However, the properties of DM particles remain elusive. Among the postulated popular DM candidates are the weakly interacting massive particles (WIMPs)~\cite{Jungman:1995df,Bergstrom:2000pn,Bertone:2004pz,Feng:2010gw}. WIMPs would naturally possess an abundance similar to today's DM when assumed to have been thermally produced in the early universe~\cite{Planck:2018vyg}: if 
the thermally-averaged DM annihilation cross section $\langle \sigma v \rangle$ is driven by velocity-independent processes, the annihilation cross section today is the same as at freeze-out in the early universe, \textit{i.e.},
$\langle \sigma v \rangle_{\rm th} \simeq$ 2.3$\times 10^{-26}$ cm$^3$s$^{-1}$~\cite{Steigman2012}. 
Such s-wave annihilation processes would produce gamma-ray signals that are within reach of \textit{Fermi}-LAT~\cite{Fermi-LAT:2016uux} and current ground-based Cherenkov telescopes~\cite{HESS:2022ygk} as well as of cosmic-ray satellite experiments~\cite{Hooper:2014ysa,Cui:2016ppb,Cholis:2019ejx,Lin:2019ljc,Ishiwata:2019aet}.

The center of the Milky Way is a prime target for indirect DM detection because of its proximity to Earth and substantial DM content.  Assuming the s-wave annihilation cross section, DM searches have been extensively performed in the Galactic Center~\cite{Abdallah:2016ygi,Abdallah:2018qtu,HESS:2022ygk} as well as nearby dwarf spheroidal galaxies~\cite{Fermi-LAT:2015att,Fermi-LAT:2016uux}
providing strong constraints on annihilation cross section for DM masses from a few ten GeV up to several ten TeV. In addition, the excess of GeV gamma rays in the inner few degrees of the GC~\cite{Hooper:2010mq,Gordon:2013vta,Fermi-LAT:2015sau,Fermi-LAT:2017opo}, known as the Galactic Center Excess (GCE), if originated dominantly from s-wave DM annihilation\footnote{An hypothetical population of millisecond pulsars has also been postulated to explain the GCE, see, for instance, Refs.~\cite{Lee:2015fea,Bartels:2015aea,Macias:2018ylj,Macias:2019omb,Pohl:2022nnd}.} may be in tension with the strong constraints obtained from the non-observation of GeV gamma rays towards nearby dwarf galaxies~\cite{Fermi-LAT:2016uux,Abazajian:2020tlm}\footnote{Note, however, that this tension could be slightly weakened in case of informative priors for their density profile parametrizations from structure formation models~\cite{Ando:2020yyk}.}.
However, in case the annihilation cross section is dependent on the relative velocity $v$ of the self-annihilating DM particles, \textit{e.g.}, for p-wave ($\sigma v \propto (v/c)^2$) or d-wave ($\sigma v \propto (v/c)^4$) annihilation, the DM annihilation rate is reduced in astrophysical environments with lower velocity dispersion as expected in dwarf galaxies compared to the GC, which therefore weakens the constraints from dSphs\footnote{An alternative is to assume that the annihilation into the electron channel dominates, in which case the GCE is produced through inverse Compton scattering which is suppressed in dSphs because of their low interstellar radiation fields, see, for instance, Ref.\cite{Calore:2014nla}.}.

More phenomenologically, plenty of DM models with symmetries impose s-wave annihilation suppression with leading contribution from p-wave annihilation. Among them are charged scalar DM annihilation through s-channel Standard Model gauge boson~\cite{Hagelin:1984wv}, 
secluded DM~\cite{Pospelov:2008jd,Tulin:2013teo,Shelton:2015aqa} and fermionic Higgs portal DM~\cite{Kim:2006af,Kim:2008pp}. A classic example of s-wave suppression is when the Majorana fermionic DM annihilate via Z boson into Standard Model fermions~\cite{Jungman:1995df}.
See, for instance, Ref.~\cite{Kumar:2013iva}, for a summary of models where the s-wave annihilation contribution is suppressed or forbidden.

Assuming a DM model with s-wave annihilation, the expected annihilation signal is proportional to the squared DM density integrated along the line of sight, usually referred to as the J-factor. In the case of p-wave annihilations, the J-factor accounts for the velocity distribution in the astrophysical object~\cite{Boddy:2018qur}. Cosmological hydrodynamical simulations such as those carried out by the Feedback In Realistic Environments (FIRE-2) collaboration (\textit{e.g.}, Refs.~\cite{Wetzel:2016wro,Hopkins:2017ycn,Lazar:2020pjs}) enable to investigate the velocity-dependent J-factors in the center of Milky Way-size galaxies. In particular, the authors of Refs.~\cite{Board:2021bwj,McKeown:2021sob} showed increased DM velocity dispersion in the central region of Milky Way-sized galaxies compared to DM-only simulations, leading to 
significant enhancements of the J-factors at a few degrees from the GC in the p and d-wave cases compared to DM-only simulations. In what follows, we will assess the constraints from current Imaging Atmospheric Cherenkov telescopes (IACTs) to p- and d-wave annihilation signals for TeV dark matter in the central region of the Milky Way. To do so, we make use of the most up-to-date H.E.S.S.-like observations of the GC with the five-telescope array~\cite{HESS:2022ygk}. These consist of the H.E.S.S. Inner Galaxy Survey~\cite{HESS:2022ygk}, which provides the currently GC region's highest exposure at very-high-energy (VHE, E $\gtrsim$ 100 GeV) gamma-rays, with a total coverage of 546 hours distributed over the inner few degrees of the Milky Way halo.

The paper is organized as follows. Section~\ref{sec:profile} describes the expected DM distribution in the GC and expected signals for velocity-dependent DM annihilation models. In Section~\ref{sec:GCgammarays}, we present the sources of VHE gamma rays in the Galactic Centre region. Section~\ref{sec:analysis} shows the analysis framework used to compute the sensitivity to velocity-dependent DM models in the GC. Results are presented in Section~\ref{sec:limits}. Section~\ref{sec:summary} is devoted to our conclusions.

\section{Velocity-dependent dark matter in the Galactic center}\label{sec:profile}
\subsection{Dark matter distribution for velocity-dependent models}
For s-wave DM annihilation models, the expected gamma-ray emission depends on the square of the DM density integrated along the line of sight. This so-called J-factor is typically computed using density profiles resulting from N-body simulations that ignore baryonic effects. However, recent realistic hydrodynamic simulations of MW-like galaxies have revealed substantial deviations 
from the spatial distribution predicted by DM-only simulations~\cite{Abazajian:2020tlm}.  In particular, some hydrodynamical simulations, including baryonic effects, predict the existence of kiloparsec-sized DM cores due to the impact of the Galactic bar and stellar feedback~\cite{Chan:2015tna}.

\begin{figure*}[!t]
\includegraphics[width=0.43\textwidth]{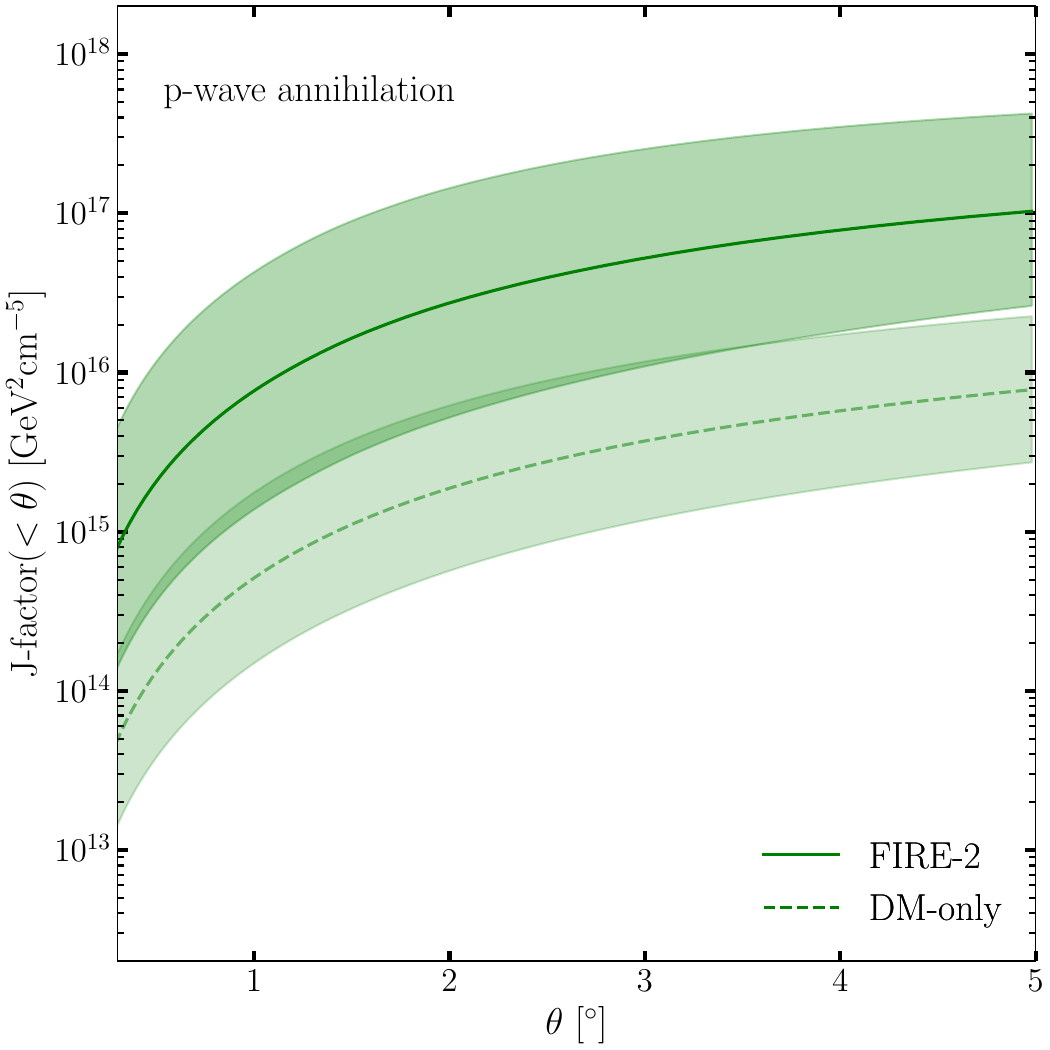}
\includegraphics[width=0.43\textwidth]{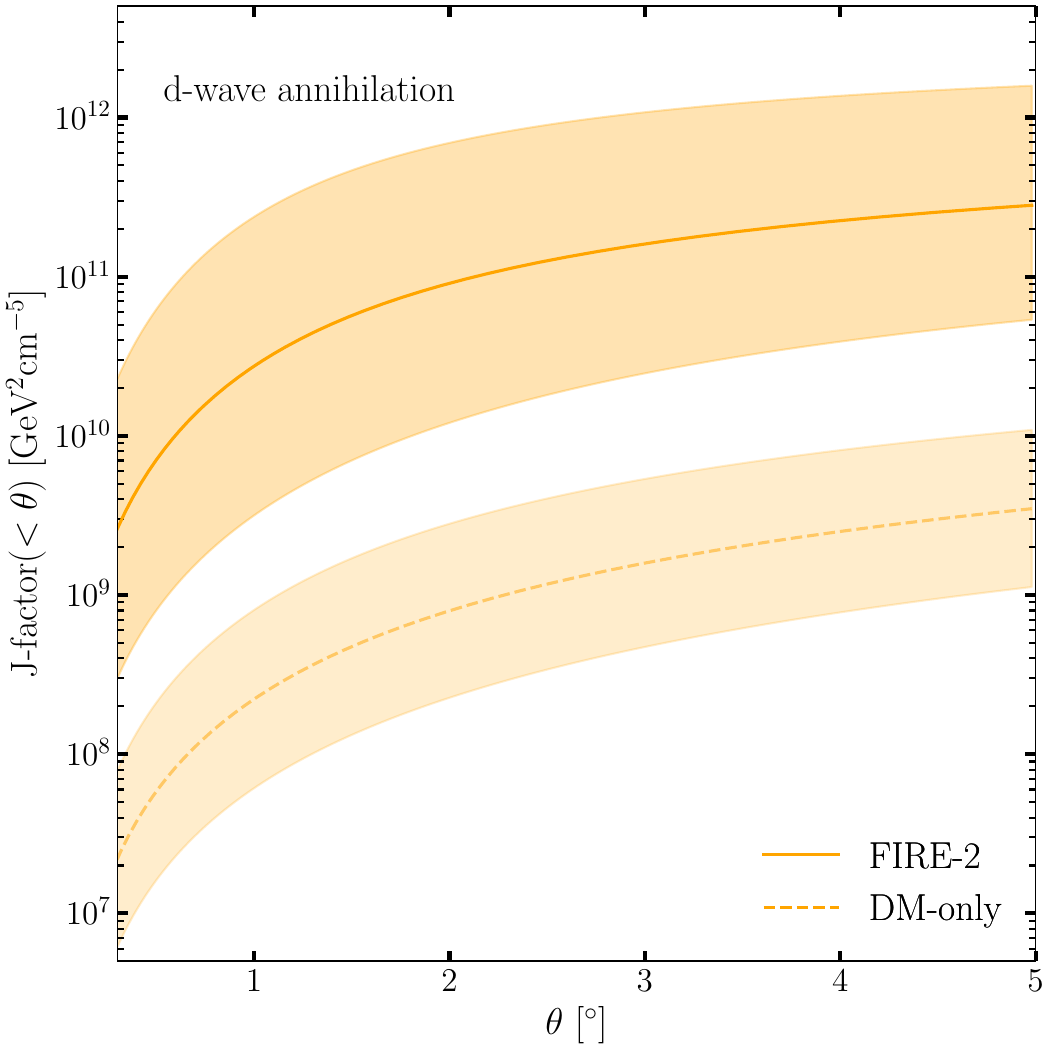}
\caption{Integrated J-factor (in GeV$^2$cm$^{-5}$) versus angular distance $\theta$ (in $^\circ$) from the Galactic Center for p-wave (left panel) and d-wave (right panel) annihilations, respectively. Both are displayed from cosmological DM-only (light-shaded area) and FIRE-2 (dark-shaded area) simulations. The mean J-factor is extracted from a set of 12 DM-only (dashed lines) and FIRE-2 (solid lines) simulations~\cite{McKeown:2021sob}. The shaded areas show the J-factor value extrema obtained in these sets.} 
\label{fig:jfactors}
\end{figure*}

For velocity-dependent models, such as the p-wave  
and d-wave 
DM scenarios, the J-factor is generalized to encompass the velocity distribution of DM particles. Reference~\cite{McKeown:2021sob} conducted various hydrodynamical simulations using FIRE-2 zoom-in simulations to compute high-resolution J-factor maps for s, p, and d-wave DM models. FIRE-2 includes radiative heating and cooling for gas, stellar feedback from OB stars, type Ia and type II supernovae, radiation pressure, and star formation effects. They found that the central DM velocity dispersion was significantly amplified in FIRE-2 (by factors of $\sim 2.5-4$) compared to DM-only simulations~\footnote{The impact of galaxy formation on the spatial distribution of DM halos are less dominant, increasing in some cases the central DM density and decreasing in others.}. Since p-wave and d-wave  DM models are strongly velocity-dependent, the FIRE-2 J-factors derived in Ref.~\cite{McKeown:2021sob} are amplified by factors of $\sim 3-60$ and $\sim 10-500$ compared to DM-only simulations, respectively. 
Figure~\ref{fig:jfactors} 
shows integrated J-factor as a function of the angular distance (from the GC) for p and d-wave annihilations, obtained from FIRE-2 and DM-only simulations, respectively.
These amplifications persist even though their DM density profiles are normalized to the same local DM density such that $\rho$(r$_\odot$ = 8.3 kpc) = 0.38 GeVcm$^{-3}$. The chosen value of the local DM density is
derived from LAMOST DR5 and Gaia DR2~\cite{2020MNRAS.495.4828G}.
The DM density at the solar location is subject to uncertainties. Following Refs.~\cite{Read:2014qva,Zyla:2020zbs}, recent 
determinations of $\rho$(r$_\odot$) are in the range (0.2 − 0.6) GeV/cm${^3}$.
As observational estimates of the local density become more precise, any change of $\rho$(r$_\odot$) can be propagated to the results by rescaling the DM signal by ($\rho(r_\odot)$/0.38 GeV/cm$^3$)$^2$.

The authors of Ref.~\cite{McKeown:2021sob} studied 12 pairs of simulations (for each FIRE-2 simulation, there is a DM-only one) suitable for comparisons with the Milky Way. 
The analysis of these zoom simulations has shown these are good candidates for comparison with the
Milky Way and used to generate synthetic surveys resembling Gaia DR2 in data  structure, magnitude limits, and observational error~\cite{2020ApJS..246....6S}. In particular, 
for each selected simulation,
the stellar mass agrees with the Milky Way one of M$_\star$ = (3 - 11)$\times$10$^{10}$ M$_\odot$.
The virial masses of the halos found in these simulations
are within the expectations fro the Milky Way 
M$_{\rm vir}$ = (0.9 - 1.8)$\times$10$^{12}$ M$_\odot$.
The averaged virial mass from the FIRE-2 simulations is in very good agreement with the value inferred by Gaia DR2 of M$^{\rm total}_{200}$=1.08$^{+0.20}_{-0.14}\times$10$^{12}$ M$_\odot$~\cite{Cautun:2019eaf}.

We have selected the minimum and maximum J-factors in the present work and computed the mean values as plotted in Fig.~\ref{fig:jfactors}, to investigate the H.E.S.S. sensitivity to velocity-dependent DM models. More details on the calculation of the integrated J-factors in the regions of interest used for the DM search are provided in Appendix~\ref{sec:appA}.

We note that the minimum spatial resolution of the FIRE-2 simulations is $\sim 400$ pc, and their study did not consider AGN-like effects from Sgr A$^\star$. In the present study, we have extrapolated the J-factor profile to the central supermassive black hole using a linear approximation. These extrapolations imply an almost 
flat density profile in the Galaxy's inner $\sim 400$ pc. This is likely a conservative approach, as it is expected that AGN feedback would cause an even more significant enhancement in DM velocities in the vicinity of the supermassive black hole 
Sgr~A$^\star$~\cite{Johnson:2019hsm,McKeown:2021sob}. 
At distance below 400 pc, Sgr A* feedback may impact DM velocities. However given the low mass of Sgr A* compared to the expectation from the black hole mass - velocity dispersion relationship found  elliptical and bulge galaxies~\cite{Kormendy:2013dxa}, no significant decrease of the velocity would be expected.

\begin{figure}[!ht]
\includegraphics[width=0.43\textwidth]{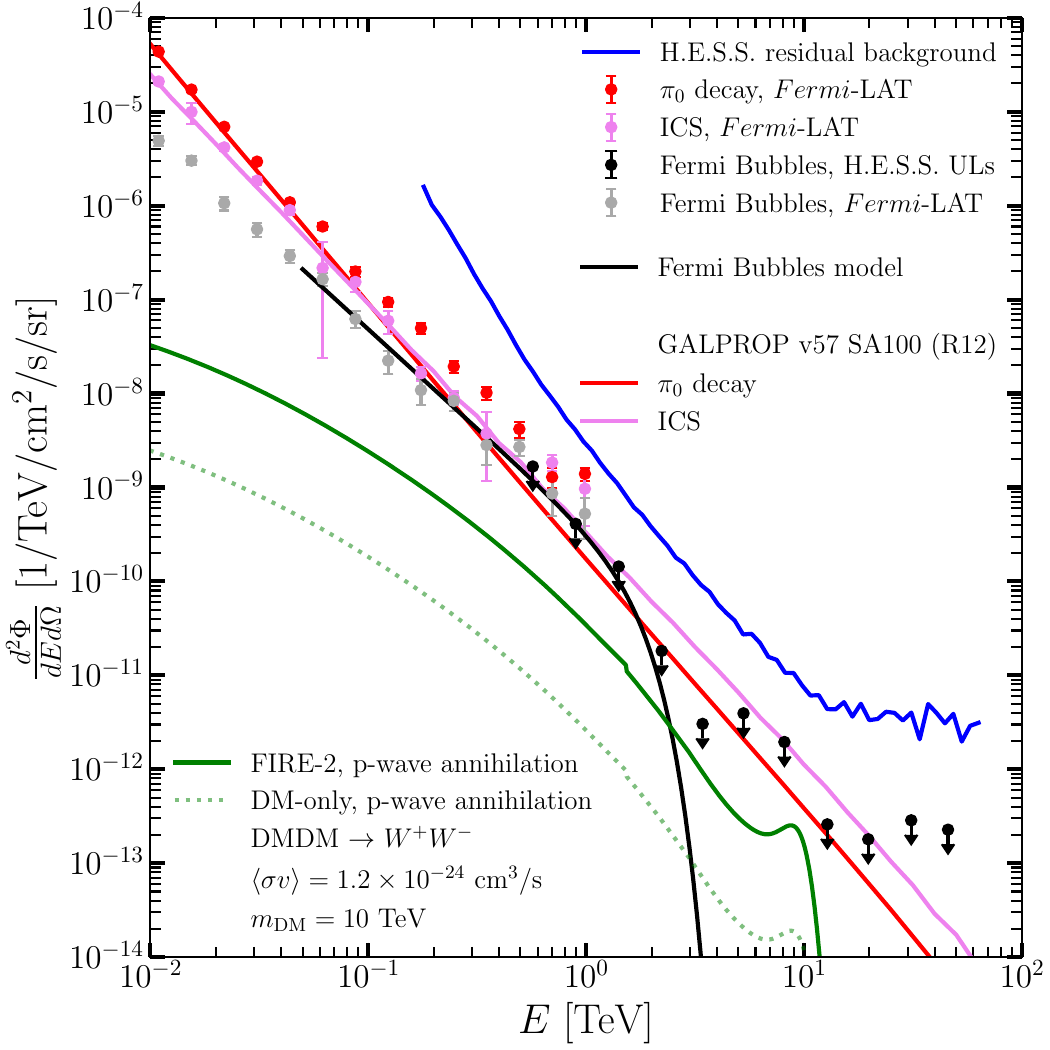}
\caption{Energy-dependent fluxes in  $d^2\Phi/dEd\Omega$ versus energy $E$ for p-wave DM annihilation signals and backgrounds in the GC region.
The $\pi_0$ (solid red line) and ICS (solid pink line) components of the Galactic diffuse emission are obtained from v57 GALPROP simulations for the 
ROI used in this work and for the model SA100 (R12). The red and pink circles show the $\pi_0$ and ICS components, respectively, as measured from \textit{Fermi}-LAT~\cite{Fermi-LAT:2017opo}. 
The residual background flux measured by H.E.S.S.~\cite{HESS:2022ygk} is shown as a solid blue line.
The flux points and upper limits from the \textit{Fermi}-LAT and H.E.S.S. analyses of the FBs~\cite{2021arXiv210810028M} are shown as grey and black points, respectively. The best-fit spectrum of the FBs emission for energies above 100 GeV is shown as the solid black line~\cite{2021arXiv210810028M}.
The expected fluxes for p-wave annihilation of DM of 10 TeV mass in the $W^+W^-$ channel with the mean DM distribution predicted from FIRE-2 (solid green line) and DM-only (dotted green line) simulations are shown for an annihilation cross section of \sigmav ~= \sigmav$_{\rm th}^{\rm p}$ = 1.2$\times$10$^{-24}$ cm$^3$s$^{-1}$.}
\label{fig:fluxes}
\end{figure}

\subsection{Expected velocity-dependent dark matter annihilation signals}\label{subsec:DMsignals}
The energy-differential flux of gamma rays produced by the pair-annihilation of Majorana DM particles of mass $m_{\DM}$ and velocities $\overrightarrow{v_1}$ and $\overrightarrow{v_2}$ in a DM halo of density $\rho$, can be expressed as\footnote{Such a formulation assumes that the DM velocity distribution $f(\overrightarrow{r},\overrightarrow{v})$ can be written as $f(\overrightarrow{r},\overrightarrow{v}) = \rho{(\overrightarrow{r})}g(\overrightarrow{v})$ normalized such that $\int d^3v f(\overrightarrow{r},\overrightarrow{v}) = \rho(\overrightarrow{r})$.}:
\begin{equation}
\begin{aligned}
\label{eq:dmflux}
\frac{d^2\Phi}{d E d\Omega}=
\frac{\langle \sigma v \rangle}{8\pi m_{\DM}^2}\sum_f  {\rm BR}_f \frac{d N_f}{d E}\, \frac{dJ_n}{d\Omega}, \\
\frac{dJ_n}{d\Omega} = \int_{\rm los}  d s\, \rho_\DM^2(r[s,\theta]) \\ 
\times \int d^3v_1 \,g_r(v_1)\int d^3v_2\, g_r(v_2) \frac{|\overrightarrow{v_1}-\overrightarrow{v_2}|^n}{c^n} \, ,
\end{aligned}
\end{equation}
where $n$ = 0, 2, 4 for s, p and d-wave annihilations, respectively. $\theta$ is the angle between the GC and the line of sight (los). The radial distance $r$ from the GC is related to the distance along the los $s$ such as $r = (s^2+r_\odot^2-2\,s\, r_\odot\,cos\,\theta)^{1/2}$, where $r_\odot$ is the distance between the GC and the Sun. 

In case of standard s-wave annhilations, the velocity-weighted cross section \sigmav\, required to match the observed DM density has been carefully computed in Ref.~\cite{Steigman2012}. Following Ref.~\cite{kolb1989,Jungman:1995df}, the thermal \sigmav\, values in case of p- and d-wave annihilations as described in Ref~\cite{McKeown:2021sob} can be estimated to 
\sigmav$_{\rm th}^p$ = 1.2$\times$10$^{-24}$ cm$^3$s$^{-1}$, and \sigmav$_{\rm th}^d$ = 4.3$\times$10$^{-23}$ cm$^3$s$^{-1}$, respectively, for $x_{\rm FO} = m_{\rm W}/T_{\rm FO}$ = 25, where  $T_{\rm FO}$ is the freeze-out temperature. 
For a broad range of masses from GeV to 100 TeV, $x_{\rm FO} = 20 - 30$, implying a 20\% and 36\% change on the thermal p- and d-wave thermal cross section, respectively.
We show the energy-dependent expected fluxes p-wave DM with a mass of 10 TeV annihilating into the $W^+W^-$ channel, 
for the mean DM distribution predicted from FIRE-2 and DM-only simulations in Fig.~\ref{fig:fluxes}. The fluxes are convolved by the finite energy resolution of the H.E.S.S. instrument, modeled as a Gaussian with width specified by $\sigma/E = 10\%$, following Ref.~\cite{2009APh32231D}.

\section{VHE gamma rays in the Galactic Center}\label{sec:GCgammarays}
\subsection{Modeling the Galactic Diffuse Emission with GALPROP}
\label{sec:GDEmodelling}

Energetic cosmic rays (CRs) interact with interstellar gas, magnetic, and radiation fields to produce diffuse gamma-ray emission, which dominates the radiation flux observed from the Galactic center in the GeV energy range and is also expected to be a very significant component at TeV-scale energies.

In this work, we use the most recent release of the CR propagation framework GALPROP (version 57)~\cite{Porter:2021tlr} to model the expected TeV-scale diffuse gamma-ray emission for a set of models that assume different realistic CR source 
density and interstellar radiation field distributions. Despite their differences, the models considered here agree with an extensive collection of locally measured CR data and are expected to be representative of the gamma-ray uncertainties related to the transport of CRs in the Galaxy.

For a given GALPROP simulation, the key inputs are the CR source density distribution, the distributions of interstellar gas, interstellar radiation fields, and magnetic fields leading to energy losses and secondary particle production. Below is a brief summary of our simulations setup.

\begin{itemize}
    \item[(i)] \textit{CR source distribution:} There are considerable uncertainties associated with the CR source distribution. We use two representative three-dimensional models that assume different injection luminosity strengths in the Galactic disc and spiral arms. Following Ref.~\cite{Porter:2017vaa,Johannesson:2018bit}, we employ the CR source distribution termed \textit{SA0}, which assumes that all sources are located in the Galactic disk, and \textit{SA100}, which places all the CR sources in the spiral arms. The first and third columns of Table I in Ref.~\cite{Porter:2017vaa} show the parameter setup assumed for models SA0 and SA100, respectively. For simplicity, we disregard the possibility of a new central source of electrons in the GC~\cite{Macias:2021boz}. This important point will be addressed in a future study.

    \item[(ii)] \textit{Interstellar gas models:} Since our CR simulations are run in three-dimensional (3D) mode, we use the 3D interstellar gas distribution models developed in Ref.~\cite{Johannesson:2018bit}. Their H$I$ and $^{12}$CO 3D maps were obtained by fitting to LAB-H$I$~\cite{Kalberla:2005ts} and CfA composite CO data~\cite{Dame:2000sp}. They provide a more accurate representation of the interstellar gas distribution than the 2D gas maps available in older versions of the GALPROP framework.

\item[(iii)] \textit{Interstellar radiation fields and magnetic field model:} Reference~\cite{Porter:2017vaa} developed 3D interstellar radiation field (ISRF) models -- namely R12 and F98 -- for the Galaxy using stellar and dust distribution functions with spatially smooth surfaces. These are based on the models introduced in Refs.~\cite{Robitaille:2012,Freudenreich:1997bx}. Despite offering equivalent solutions for the ISRF intensity distribution in the Galaxy, neither matches the data perfectly. As a result, we run GALPROP simulations with both ISRF models to estimate the extent of the modeling uncertainty associated with our incomplete knowledge of the distribution of the ISRF.

We use the magnetic field model of  Ref.~\cite{Pshirkov:2011} (available in GALPROP) for the synchrotron energy losses. We are considering an exponential disk model with a strength of  $5\;\mu$G at the solar position.

\item[(iv)] \textit{Spatial resolution of the simulations:} We perform the GALPROP runs using a non-linear (tangential) transformation of coordinates~\cite{Porter:2021tlr} allowing us to achieve 50 pc spatial resolution in the GC, and 1 kpc in the outskirts of the Galaxy. Each simulation required  $\sim 290$ Gb of memory and $\sim 8$ h of running time on a 128-core computing node.
\end{itemize}

The spectral and morphological behaviors and the flux overall normalization of $\pi_0$ and ICS components are provided by the GALPROP predictions. We take these normalizations to match the measured values from \textit{Fermi}-LAT~\cite{Fermi-LAT:2017opo}, to input a physical modeling from GeV to TeV energies. The energy-dependent fluxes for the $\pi_0$ and ICS components, for the model SA100 (R12) adopted for the GDE predictions with GALPROP, are shown in Fig.~\ref{fig:fluxes}. The measured $\pi_0$ and ICS energy-differential fluxes measured with \textit{Fermi}-LAT~\cite{Fermi-LAT:2017opo} are shown in the same figure.

The energy-differential fluxes for the $\pi_0$ and ICS components for all the other models considered for the GDE -- SA0 (R12), SA100 (F98) and SA0 (F98) -- are shown in Fig.~\ref{fig:fluxes_allgalprop} in Appendix~\ref{sec:appB}.
Figure~\ref{fig:hadronicMap} in Appendix~\ref{sec:appB} shows the Galactic diffuse emission model SA100 (R12) obtained from the sum of the $\pi_0$ and ICS components at an energy of 10 TeV. 
The main results obtained in this work consider SA100 (R12) as a baseline model for GDE predictions.

\begin{figure*}[ht!]
    \centering
    \includegraphics[width=0.9\textwidth]{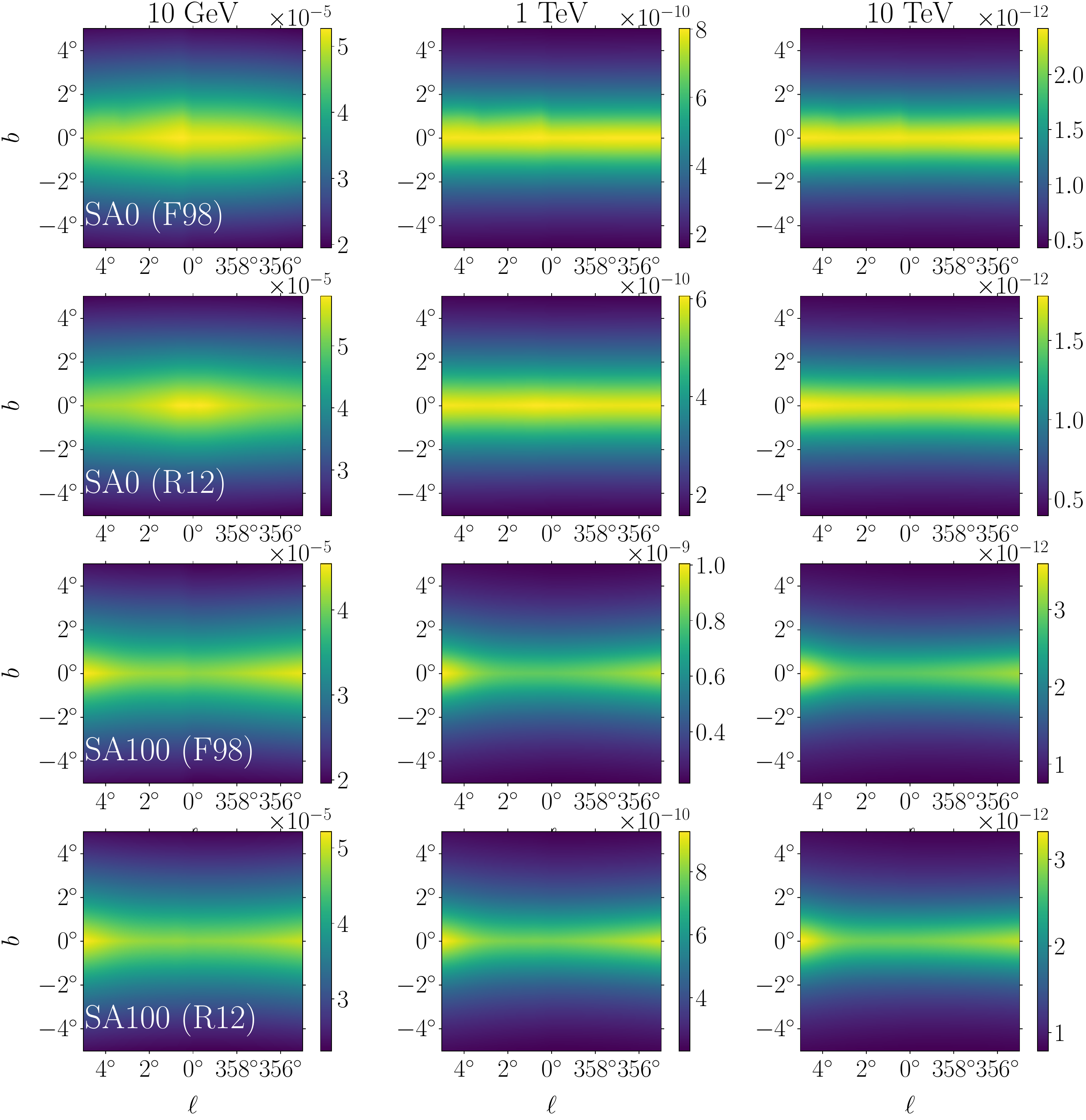}
    \caption{Inverse Compton flux maps ($d^2\Phi/dE~d\Omega$) in Galactic coordinates ($l$,$b$) predicted by GALPROP v57~\cite{Porter:2021tlr} (in units of TeV$^{-1}$ cm$^{-2}$ s$^{-1}$ sr$^{-1}$). Each row shows the IC morphology for a particular choice of CR sources density (SA0 or SA100) and ISRF model (F98 or R12). The columns display the maps at three different energies: 10 GeV, 1 TeV, and 10 TeV, respectively. The propagation parameter setup used in this simulation agrees with local CR measurements. See text for more details.} 
    \label{fig:ICSmaps}
\end{figure*}

\begin{figure*}[ht!]
     \centering
   \includegraphics[scale=0.34]{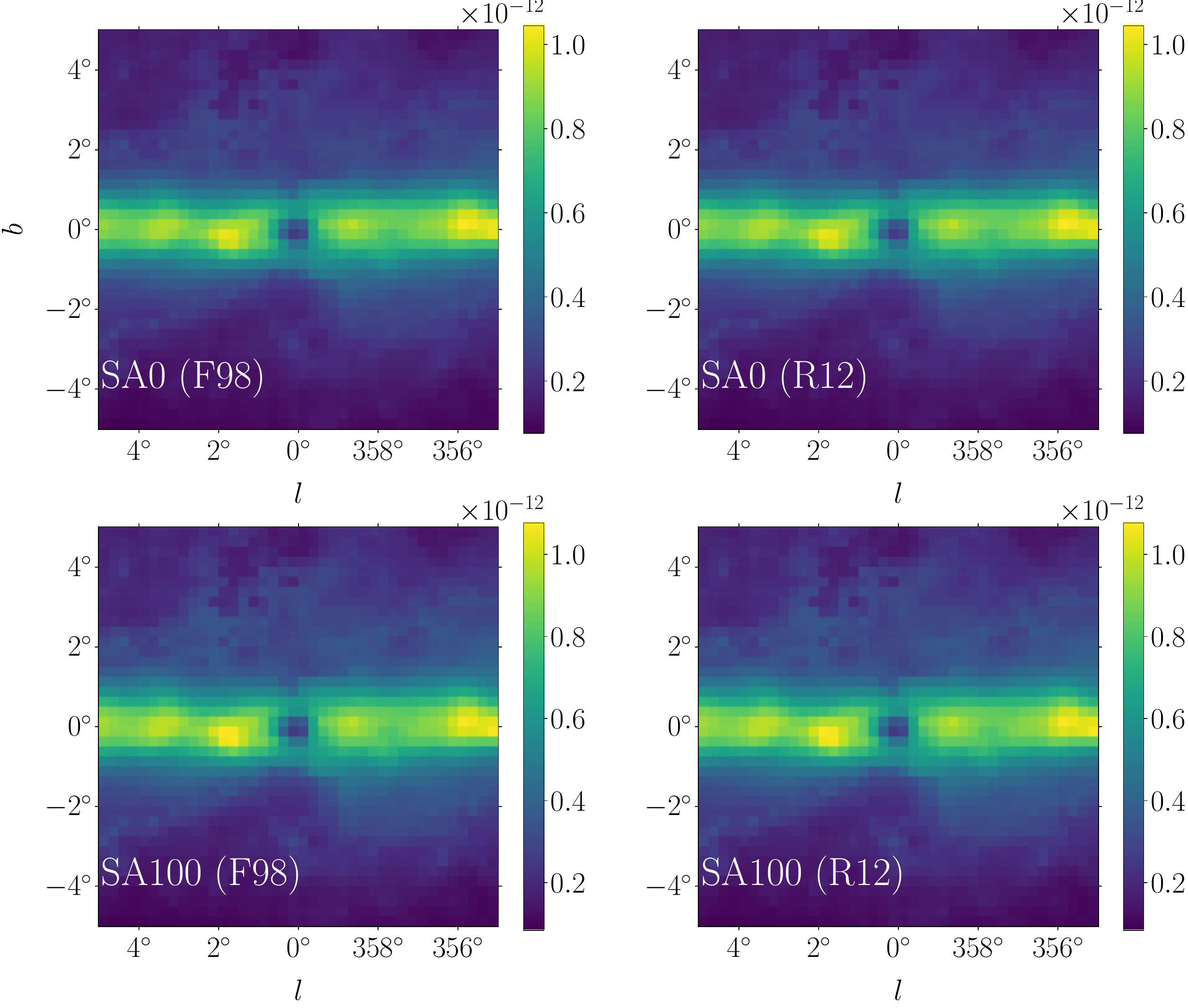}
     \caption{Hadronic flux maps ($d^2\Phi/dE~d\Omega$) in Galactic coordinates ($l,b$) at an energy of 10 TeV computed with GALPROP v57~\cite{Porter:2021tlr} (in units of TeV$^{-1}$cm$^{-2}$s$^{-1}$sr$^{-1}$). Each panel shows the predicted morphology for a particular choice of CR source density (SA0 or SA100) and ISRF model (F98 or R12). Note that all models agree with local CR measurements by construction and that the spatial morphology of the maps is heavily dominated by the distribution of interstellar gas.}
     \label{fig:hadronicmaps}
 \end{figure*}

The flux maps for the simulated IC component are presented in Fig.~\ref{fig:ICSmaps}. Each row corresponds to a CR source density (SA0 or SA100) and an ISRF model (F98 or R12). The three different columns show the spatial morphologies at 10 GeV, 1 TeV and 10 TeV, respectively. These are determined by a convolution of the CR source distribution and the geometry of the ISRF components. As the energy of the CR $e^{\pm}$ increases, the Klein-Nishina effects become important, and the spatial morphology of the IC map resembles more and more the spatial distribution of the CR sources. This is because, in this case, the relevant target photons are the Cosmic Microwave Background photons, which are isotropically distributed.
Figure~\ref{fig:hadronicmaps} shows the flux maps for the hadronic component at an energy of 10 TeV for the CR source densities (SA0 or SA100) and the ISRF models (F98 or R12) considered in this work. As expected, the spatial morphology is mostly driven by the target material density.

\subsection{Backgrounds in the central region on the Milky Way}\label{subsec:HESS-data}
Any analysis searching for DM annihilation signals with IACT observations of the GC region has to deal with the irreducible gamma-ray background, which comes from conventional astrophysical emissions arising from a variety of objects and production processes, as well as from hadron and electron cosmic rays (CRs) that are misidentified as gamma rays due to the finite discrimination between gamma-rays and CRs~\cite{BERNLOHR2013171}. The latter is referred to hereafter as the residual background.
Its determination in the GC  and blank-field regions may differ due to the different levels of night sky background in these environments. However, the method used in Ref.~\cite{HESS:2022ygk} can properly determine the background in the signal region from regions with different night sky background conditions. {It is, therefore, safely assumed that the residual background measurement performed in Ref.~\cite{HESS:2022ygk} applies to extragalactic blank-field observations.
For the analysis, we adopt the ON and OFF method. The region of interest (ROI) for DM search is assumed as the ON region. The OFF region provides the control dataset to constrain the dominant CR background.

The observations of the GC with H.E.S.S. revealed a complex astrophysical region with the detection of the central source HESS~J1745-290~\cite{Aharonian:2004wa,Aharonian:2009zk} coincident with the supermassive black hole Sagittarius A*, as well as diffuse emission connected to the interaction of relativistic particles in clouds of the central molecular zone~\cite{Aharonian:2006au}, and more recently within the inner 50 pc of the GC~\cite{Abramowski:2016mir} from PeV protons interacting in the interstellar medium. To avoid complex modeling of these background emissions and leakage of VHE emissions in the ROI for DM search, a conservative set of exclusion regions is considered here following Ref.~\cite{HESS:2022ygk}. The survey carried out by the H.E.S.S. instrument provides an accurate measurement of the residual CR background of the GC region~\cite{HESS:2022ygk}. The energy-differential flux of the residual background is shown in Fig.~\ref{fig:fluxes} and is used as a baseline for the residual VHE gamma-ray background affecting DM signal searches with IACTs observations.

Another potentially significant background emission in the GC region comes from the Fermi Bubbles (FBs), the giant double-lobe emission detected using the \textit{Fermi}-LAT satellite \cite{2010ApJ...724.1044S}. We consider here the flux points and upper limits from the \textit{Fermi}-LAT and H.E.S.S. analyses given in \cite{2021arXiv210810028M}. We also extract the best-fit spectrum for energies above 100 GeV from the same reference. However, as demonstrated in Refs.~\cite{Rinchiuso:2020skh,Montanari:2022buj}, subtracting the FBs emission from the overall background for DM searches is negligible on the sensitivity reach. The \textit{Fermi}-LAT and H.E.S.S. flux points and upper limits, together with the best-fit spectrum are shown in Fig.~\ref{fig:fluxes}.

The most intense astrophysical background in the GC region is generated by energetic cosmic rays interacting with interstellar gas, magnetic, and radiation fields, as described in Sec.~\ref{sec:GDEmodelling}.

Several studies~\cite{Linden:2016rcf,Calore:2015bsx,Macias:2018ylj,Abazajian:2020tlm,Pohl:2022nnd} have discussed the possibility of a high-energy tail in the Galactic Center excess (GCE), which might extend up to a few TeV. If the GCE is due to a new population of millisecond pulsars, then this high-energy tail could naturally arise from the injection of TeV-scale electrons/positrons from the putative population of MSPs (\textit{e.g.}, Ref.~\cite{Macias:2021boz}). 
However,  we do not model the high-energy tail in our analysis in order to compute conservative upper limits for p- and d-wave DM.

\begin{figure*}[!ht]
\includegraphics[width=0.44\textwidth]{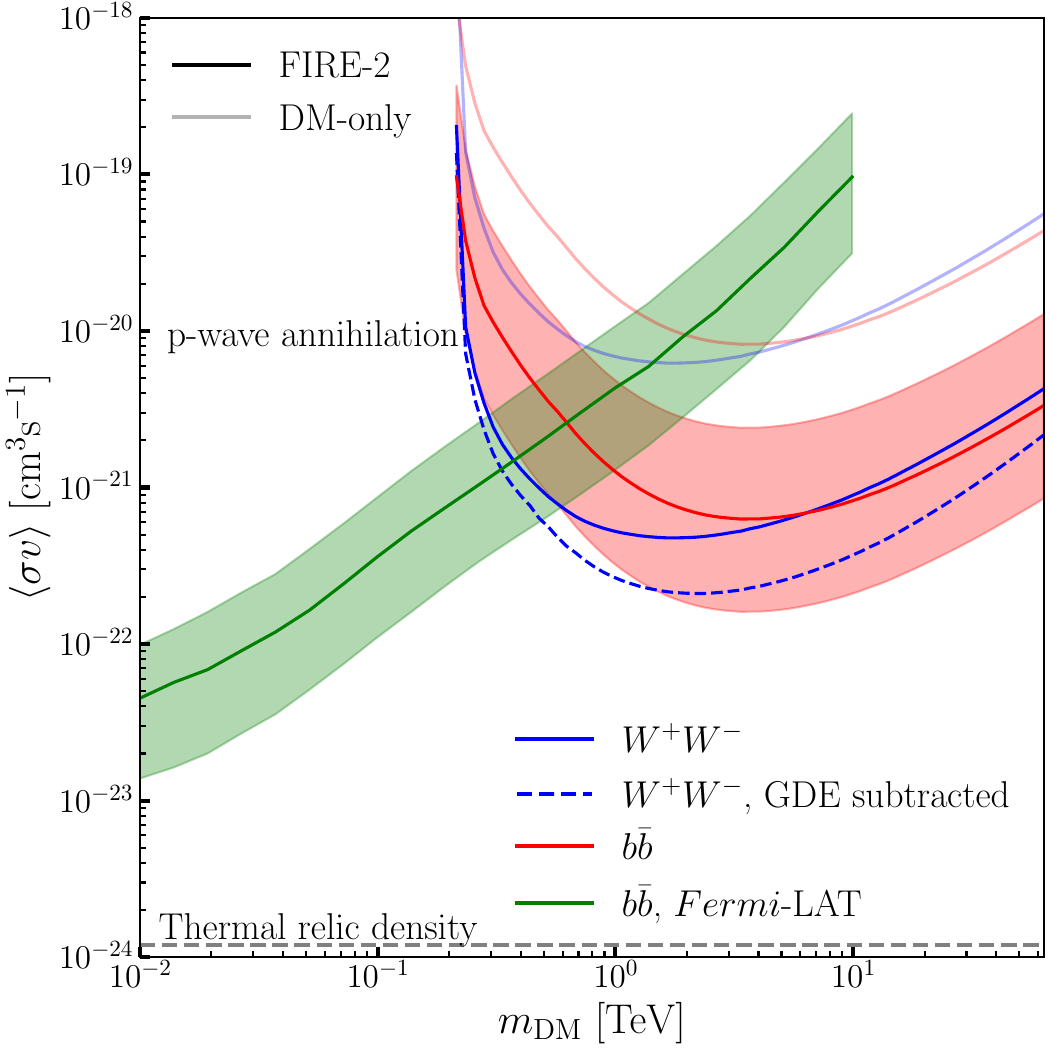}
\includegraphics[width=0.44\textwidth]{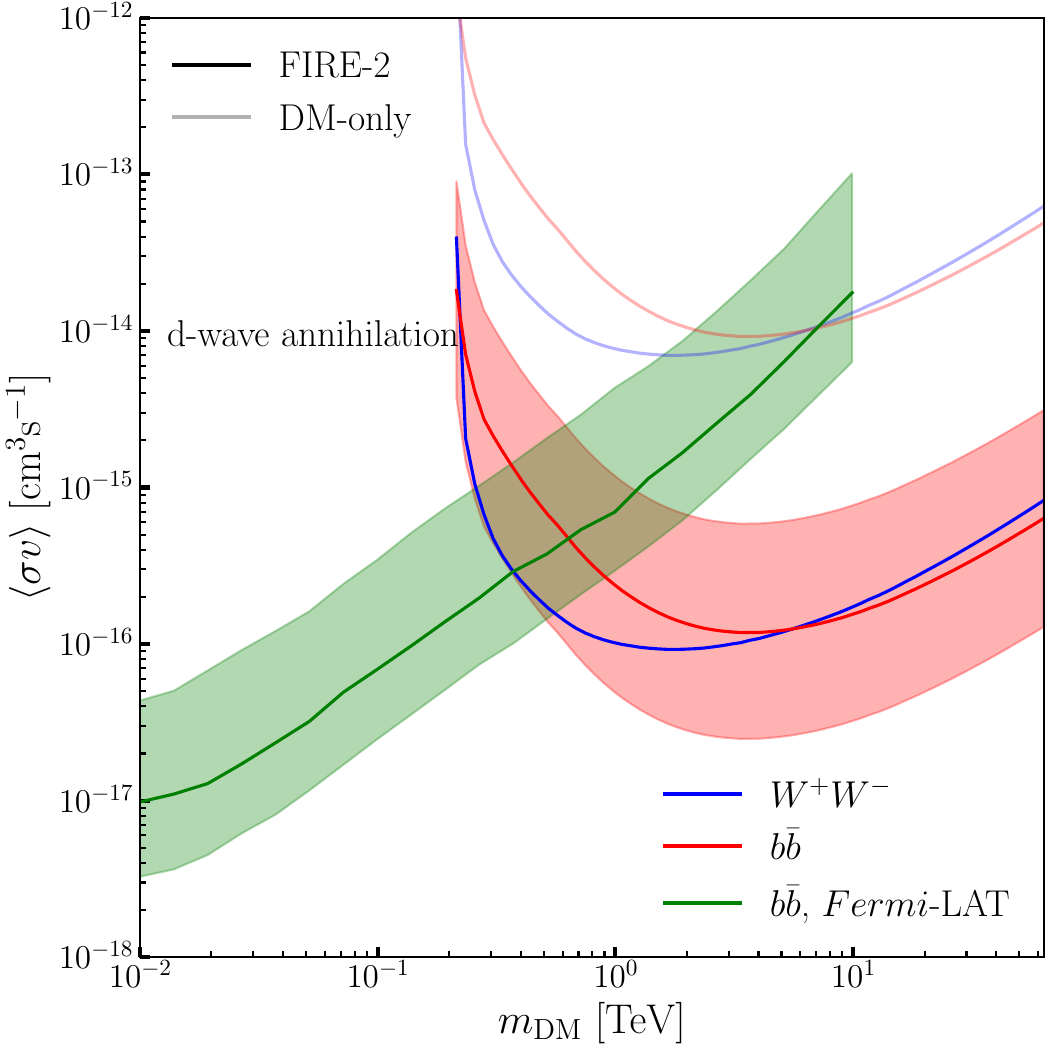}
\caption{95\% C.L. mean expected upper limits on \sigmav~versus the DM mass, assuming p-wave (left panel) and  d-wave (righ panel) annihilations, respectively, in the $W^+W^-$ (solid blue line) and $b\bar{b}$ (solid red line) channels, respectively. The limits obtained when the GDE is subtracted from the measured residual background are shown for the $W^+W^-$ channel (dashed blue line). The dark and light-shaded lines correspond to the FIRE-2 and DM-only simulations, respectively. The horizontal grey dashed line corresponds to the expected thermal annihilation cross section for the p-wave annihilation signal. The \textit{Fermi}-LAT constraints are shown as a solid green line~\cite{McKeown:2021sob}. The containment bands show the limits obtained when considering the maximum and minimum values of the J-factors from the FIRE-2 simulations used in this analysis.}
\label{fig:Limits}
\end{figure*}

\section{Statistical analysis}
\label{sec:analysis}
The statistical analysis and the derivation of upper limits is performed following the well-established procedure based on the definition of a 2-dimensional log-likelihood ratio test statistic (see, for instance, Refs.\cite{Silverwood_2015, PhysRevD.91.122003, Moulin:2019oyc, HESS:2022ygk, Montanari:2022buj}). This exploits the DM signal's expected spectral and spatial features in logarithmically-spaced energy bins, defined for an energy range spanning between 200 GeV and 70 TeV, and spatial bins corresponding to the ROI, respectively. The binned two-dimensional likelihood function for a given DM mass reads as:
\begin{equation}\begin{aligned}
    &\mathcal{L}_{ij} (N_{ij}^S, N_{ij}^B, \bar{N}_{ij}^S, \bar{N}_{ij}^B\, |\, N_{ij}^{\rm ON}, N_{ij}^{\rm OFF}) \\ 
    =\,\, &\textrm{Pois}[(N^S_{,ij} + N^B_{ij}), N^{\rm ON}_{ij}] \\
    \times\,\, &\textrm{Pois}[(\bar{N}^S_{ij} + \bar{N}^B_{ij}), N^{\rm OFF}_{ij}] 
    \label{eq:likelihood}
\end{aligned}\end{equation}
The number of measured events in the ON and OFF regions, in the spectral bin $i$ and the spatial bin $j$, is given by $N^{\rm ON}_{ij}$ and $N^{\rm OFF}_{ij}$. They are obtained with the procedure explained in Sec.~\ref{subsec:HESS-data}. The expected numbers of events in the signal and background regions are expressed by $N^S_{ij} + N^B_{ij}$ and $\bar{N}^S_{ij} + \bar{N}^B_{ij}$, respectively. The background is assumed to be measured from extragalactic blank-field observations such that $\bar{N}^S_{ij}$ can be considered negligible compared to $N^S_{ij}$. The expected number of background events in the $(i,j)$ bin for the ON and OFF regions is given by $N^B_{\rm ij}$. 
For the computation of the sensitivity, the background model, referred as to residual background, has no free parameter since it is directly extracted from measurement. The residual background taken from the OFF region is assumed to be a perfect description on the background model in the ON region.
Our background-modeling approach with real data would require the consideration of possible systematic uncertainties on the expected residual background in the ON region from the OFF measurement (see Ref.~\cite{HESS:2022ygk} for more details).
For the GDE, we do not allow its normalization, for example, to vary.
Instead, the impact of the mismodelling of the GDE component on the sensitivity is studied via the use different representative models of the GDE presented in Sec.~\ref{sec:GDEmodelling}.
The ON-OFF approach we use here has the advantage of factoring out instrumental systematic uncertainties  which should affect in a similar way the background in the OFF and ON regions. However, it requires to define OFF region with similar background to the ON region, which may be challenging due to the non-uniform GDE spatial morphology. We do not aim here to define the optimal observational pointing position strategy and the definition of the OFF regions. We use here a simplified approach where we model the expected background in the ON region, and our results correspond to a best-case scenario to velocity-dependent DM signals.
The number of expected events from DM annihilations, $N^{\rm S}$,
is obtained by folding the expected DM flux, given in Eq.~(\ref{eq:dmflux}) with the energy-dependent acceptance and energy resolution of the H.E.S.S. instrument, extracted from Ref.~\cite{HESS:2022ygk}. The gamma-ray yield $dN^f_\gamma/dE_\gamma$ in the channel $f$, is obtained from 
\texttt{PPPC4DMID} computational tools from Ref.~\cite{Cirelli:2010xx}. $N^S_{\rm ij}$ is computed for the different DM profiles we consider in the analysis. The J-factor values obtained for each ROI are reported in Tab.~\ref{tab:Jfactab} for the FIRE-2 and DM-only simulations, respectively, in the case of p and d-wave annihilations. Once the DM mass is chosen, the only free parameter is the strength of signal controlled by $\langle\sigma v\rangle$.

Upper limits on \sigmav~are obtained using the test statistics (TS) defined with the full likelihood function. The latter is the result of the product of the binned two-dimensional Poisson likelihood function $\mathcal{L}_{ij}$ over the spectral (\textit{i}) and spatial (\textit{j}) bins. Therefore, the TS depends on the particle DM properties, which for a given spectrum, are specified by $\langle \sigma v \rangle$ and $m_\DM$. For a given DM mass and annihilation channel, the TS can be defined as
\begin{equation}
    \text{TS}(m_{\DM}) = - 2 \log \frac{\mathcal{L}(\langle \sigma v \rangle, m_{\DM})}{\mathcal{L}(\widehat{\langle \sigma v \rangle}, m_{\DM})} \, .
    \label{eq:TS}
\end{equation}
The value of the cross section maximizing the likelihood for a given $m_{\DM}$ is given by $\widehat{\langle \sigma v \rangle}$. When the limit of large statistics is reached, the TS follows a $\chi^2$ distribution with a single degree of freedom $\langle \sigma v \rangle$. With this assumption, we compute one-sided 95\% confidence level (C.L.) upper limits on $\langle \sigma v \rangle$ extracting the cross section value corresponding to ${\rm TS} = 2.71$~\cite{Cowan:2011an}. The upper limits shown in this work are derived as the mean expectation by applying the above procedure to the Asimov dataset, for which the measured background is considered as the data to compute the mean of the expected sensitivity~\cite{Cowan:2011an}. This procedure is an accurate estimate when compared to computations with a large set of Monte Carlo datasets, as shown in Ref.~\cite{Montanari:2022buj}. In what follows, the Asimov procedure is also used to compute confidence intervals of the expected sensitivity~\cite{Cowan:2011an}.

\section{Results}
\label{sec:limits}
We search for DM annihilation signals in an ROI defined as a disk of radius up to $4^\circ$. Following Ref.~\cite{HESS:2022ygk}, the ROI is further divided into concentric annuli of width $\Delta \theta = 0.1^\circ$, to exploit the spatial characteristics of the DM signal in contrast to the background. The ROI is here extended such that the annuli's inner radii range from $\theta_i = 0.5^{\circ}$ up to $\theta_i = 3.9^{\circ}$. The exposure in each annulus considered in this analysis is extracted from Ref.~\cite{HESS:2022ygk} and applied to compute the measured number of events from irreducible VHE gamma-ray background using the residual fluxes shown in Fig.~\ref{fig:fluxes}. 

The left panel of Fig.~\ref{fig:Limits} shows the 95\% C.L. mean expected upper limits on the velocity-weighted cross section for p-wave Majorana WIMPs annihilating in the $W^+W^-$ and $b\bar{b}$ channels, respectively, for the above-mentioned velocity-dependent J-factors in case of FIRE-2 and DM-only simulations. The results including the confidence intervals obtained with the Asimov procedure are provided in Appendix~\ref{sec:appC}.

The limits reach 4.6$\times$10$^{-22}$ cm$^3$s$^{- 1}$ and $7.8\times$10$^{-22}$ cm$^3$s$^{− 1}$ for a DM particle mass of 1.7 TeV in the $W^+W^−$ and $b\bar{b}$ annihilation channels, respectively. For a DM mass of about 1 TeV, considering annihilation in the $b\bar{b}$ channel, our limits improve upon the limits derived from \textit{Fermi}-LAT data extracted from Ref.~\cite{McKeown:2021sob}, by a factor $\sim$4. Accurate modeling of the expected Galactic diffuse emission in the TeV range would allow for an improvement in the limits by a factor $\sim$2.5 for a DM particle mass of 1.7 TeV. This is shown in Fig.~\ref{fig:Limits}, for the limits obtained when the baseline GDE model is subtracted from the H.E.S.S. measured residual background.
The right panel of of Fig.~\ref{fig:Limits}
show limits for d-wave annihilation, using J-factors from FIRE-2 and DM-only simulations and considering annihilation into the same channels as in the right panel of the same figure. The limits reach 9.2$\times$10$^{-17}$ cm$^3$s$^{- 1}$ and $1.5\times$10$^{-16}$ cm$^3$s$^{− 1}$ for a DM particle mass of 1.7 TeV in the $W^+W^−$ and $b\bar{b}$ annihilation channels, respectively.

\begin{figure}[!ht]
\includegraphics[width=0.44\textwidth]{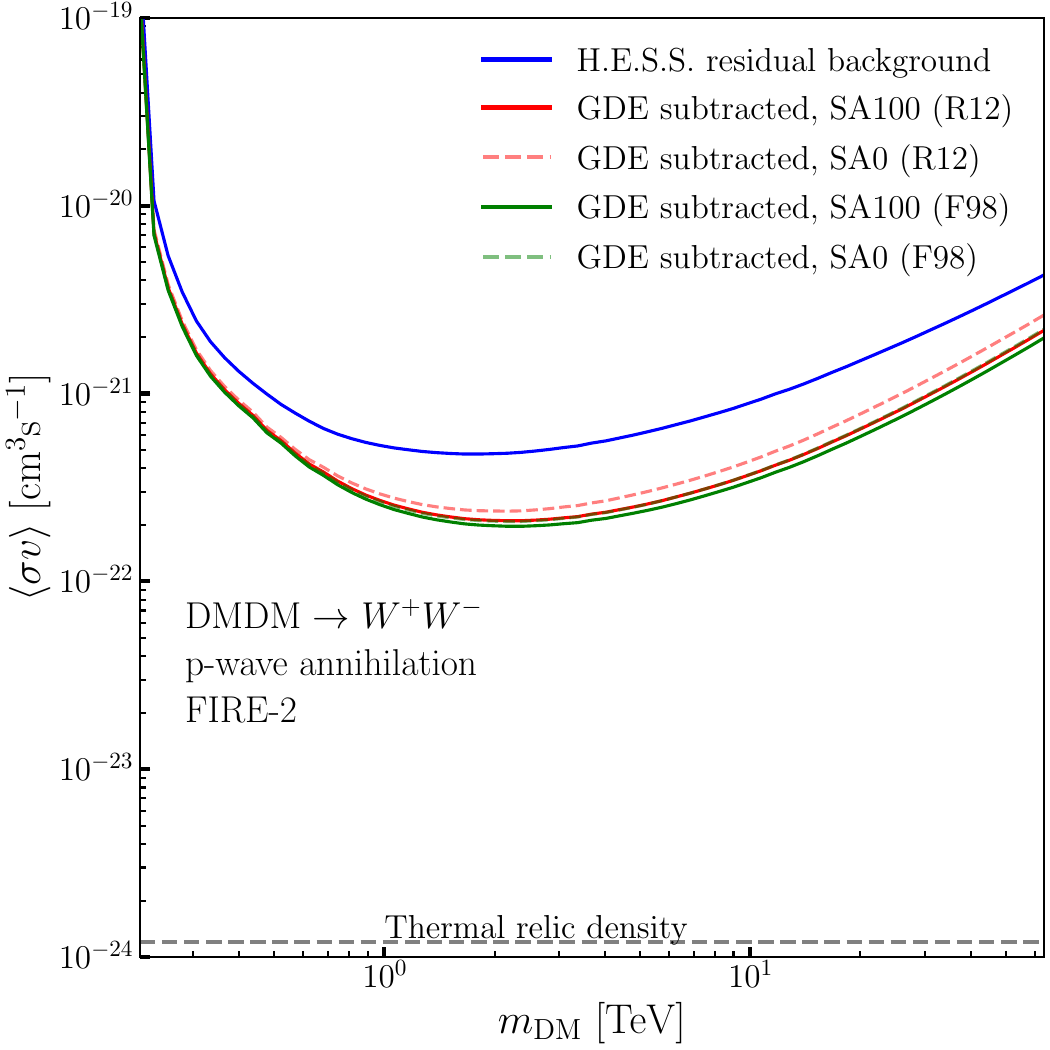}
\caption{95\% C.L. mean expected upper limits on \sigmav~versus the DM mass, assuming p-wave annihilation in the $W^+W^-$ channel. The limits obtained when the GDE is subtracted from the measured residual background are shown for all the models tested in this work -- SA100 (R12), SA0 (R12), SA100 (F98), and SA0 (F98) as solid red, dashed red, solid green and dashed green lines, respectively. The limits when no GDE modeling is subtracted are shown as the solid blue line.}
\label{fig:LimitsCompGDE}
\end{figure}

As presented in Sec.~\ref{sec:profile}, we have extracted the minimum and maximum J-factors for FIRE-2 and DM-only simulations. Fig.~\ref{fig:Limits} shows how the limits obtained for p- and d-wave annihilations into the $\bar{b}b$ channel would improve (degrade) when the maximum (minimum) J-factors are considered. This manifests that a non-negligible uncertainty for the signal prediction is obtained from the choice of the J-factor. For a DM mass of 1.7 TeV, adopting the maximum (minimum) J-factor would result in limits a factor $\sim$3.8 more (less) constraining.

Fig.~\ref{fig:Limits} showed the improvement obtained on the constraints when the GDE is modeled with the baseline GDE model
adopted in this work. 
As mentioned above, the GDE background model has no free parameter. Instead, we show the limits obtained when modeling the GDE with the other setups, as discussed in Sec.~\ref{subsec:HESS-data}, and subtracting it from the H.E.S.S. measured residual background in Fig.~\ref{fig:LimitsCompGDE}, in order to quantify the impact of the GDE mismodelling. The improvement when the diffuse emission is modeled following the SA0 (R12), SA100 (F98) and SA0 (F98) CR source density (ISRF) models is shown and compared to the previously shown limits with no GDE subtraction and with modeling according to the SA100 (R12) setup. With SA0 (R12), SA100 (F98) and SA0 (F98), the limits would improve by factors $\sim$2.2, $\sim$2.7 and $\sim$2.5, respectively, for a DM particle mass of 1.7 TeV.

\section{Discussion and Summary}
\label{sec:summary}
In this work, we have presented constraints on velocity-dependent DM models based on the latest H.E.S.S. deep VHE observations of the central region of the Milky Way, where the highest velocity-dependent DM annihilation signals are expected. To model the DM distribution in the GC region, we use the state-of-the-art FIRE-2 zoom hydrodynamical simulations of Milky Way-like galaxies. We have determined that for p-wave DM particles of about 1.7 TeV (annihilating into the $W^+W^−$ channel), the velocity-weighted annihilation cross section has an upper limit of $4.6 \times 10^{-22}$ cm$^3$s$^{-1}$ at 95\% C.L. These constraints are the strongest so far in the TeV mass range, improving limits obtained with \textit{Fermi}-LAT data above 250 GeV~\cite{McKeown:2021sob}. For a DM mass $m_{\rm DM} = 1$ TeV, our limits for annihilation into the $b\bar{b}$ channel improve \textit{Fermi}-LAT results by a factor $\sim$4.
Alternative searches for velocity-dependent DM models focus on the local largest bounded structures where the highest velocity dispersions are measured. Compared to galaxy cluster searches~\cite{Kostic:2023arx}, our results improve by about two orders of magnitude for a DM mass of 500 GeV.

Further, we have investigated to some extent the impact of Galactic diffuse gamma-ray emission mismodeling on our constraints by using different configuration sets for the GDE predictions with GALPROP. Our findings indicate that this form of uncertainty impact the constraints by about 15\% on average over the considered mass range as shown in Fig.~\ref{fig:LimitsCompGDE}.

Our study utilized the DM velocity-dispersion and spatial distribution as predicted by Ref.~\cite{McKeown:2021sob}, which align with those independently obtained by Ref.~\cite{Board:2021bwj}. In both hydrodynamic simulations, DM particle dispersion velocities are amplified towards the GC. This results in higher J-factors for velocity-dependent annihilation than DM-only simulations. We chose to solely rely on the J-factors determined by Ref.~\cite{McKeown:2021sob} due to their superior spatial resolution. Additionally, Ref.~\cite{Board:2021bwj} limits their analysis region to areas beyond 10 degrees of the GC.

Unresolved VHE sources in the GC region, \textit{i.e.}, individual sources below the detection threshold, could also contribute to the total Galactic diffuse gamma-ray emission, providing an additional contribution to the truly diffuse emission measured by H.E.S.S. A recent study has investigated the unresolved source population contribution from the H.E.S.S. Galactic plane survey and the detection potential of CTA~\cite{Marinos:2022tdj}. We note that if we subtracted this unresolved source component from the GC region, our p-wave annihilating DM constraints could be further improved. We leave this interesting possibility for future work.

The H.E.S.S. observatory offers exceptional sensitivity to the Galactic Center, compared to other IACTs in operation. Data collected by H.E.S.S. offer unprecedented insight into annihilating TeV-scale DM particles in regions of the sky where the highest dispersion velocities are expected at Galactic scales. This results in significant velocity dependence on the annihilation cross-section. 

\section*{Acknowledgments} 
We are grateful to  Shunsaku Horuichi and James Bullock for providing digitized data of Fig. 7  of Ref.~\cite{McKeown:2021sob}. We also thank Shunsaku Horiuchi and Satoshi Shirai for helpful discussions at the early stages of the project. OM was supported by a GRAPPA Prize Fellowship. In conducting this research, we utilized the GALPROP (v57) framework for cosmic ray propagation. We are grateful to the GALPROP developers for their invaluable support.

\appendix
\section{J-factors for p- and d-wave annihilation models}
\label{sec:appA}
The integrated J-factors (GeV$^2$cm$^{-5}$), as a function of the angular distance $\theta$ from the GC (in $^\circ$), are obtained for p- and d-wave  annihilation from FIRE-2 and DM-only simulations~\cite{McKeown:2021sob}, as shown in the left and right panels of Fig.~\ref{fig:jfactors}, respectively.

The distance between the GC and the Sun is taken at $r_\odot$ = 8.3 kpc. The value of the dark matter density 
at the solar location is taken as $\rho(r_\odot)$ = 0.38 GeV$^2$cm$^{-5}$. 
A significant enhancement in p- and d-wave J-factors in hydrodynamic simulations compared to DM-only cases is obtained. In the inner 5$^\circ$ of the Galactic Centre, the J-factors are enhanced by a factor $\sim$14 and $\sim$96 for the p- and d-wave annihilations, respectively. The containment bands represent the extrema of the J-factor values obtained from the set of the 12 FIRE-2 simulations~\cite{McKeown:2021sob}.

Table~\ref{tab:Jfactab} gives the J-factor values integrated in each ROI adopted in this work for p- and d-wave annihilations in the case of the FIRE-2 and DM-only cosmological simulations.
{\footnotesize
\renewcommand{\arraystretch}{1.5}
\setlength{\tabcolsep}{10.5pt}
\setlength{\arrayrulewidth}{1.3pt}
\begin{table*}[ht!]
\centering
	\begin{center}
		\begin{tabular}{cccccc}
                \hline
                \hline
			  \multirow{2}{*}{$\begin{array}{c}\text{$i^{\rm th}$ ROI }\end{array}$}&  \multirow{2}{*}{$\begin{array}{c}\text{Solid angle:}\\[-5pt] \Delta\Omega_i\,\,\, \big[10^{-4} \text{ sr}\big]\end{array}$}  & 
      \multicolumn{4}{c}{J-factor: J($\Delta\Omega_i$)  }\\
      \cline{3-6}
                & & \multicolumn{2}{c}{p-wave [10$^{15}$ GeV$^2$cm$^{-5}$sr]} & \multicolumn{2}{c}{d-wave [10$^{9}$ GeV$^2$cm$^{-5}$sr]}  \\ 
                &  &  DM-only & FIRE-2  
                & DM-only & FIRE-2  \\
			\hline
                1 & 1.05 & 0.22 & 3.28 & 0.09 & 11.82 \\
                
			2 & 1.24 & 0.26 & 3.84 & 0.11 & 13.99 \\
   
			3 & 1.44 & 0.30 & 4.40 & 0.13 & 16.09 \\
   
			4 & 1.63 & 0.34 & 4.91 & 0.14 & 17.90 \\
   
			5 & 1.82 & 0.37 & 5.43 & 0.16 & 19.62 \\

                6 & 2.01 & 0.41 & 5.92 & 0.17 & 21.08 \\
			
			7 & 2.20 & 0.44 & 6.42 & 0.19 & 22.38 \\
			
			8 & 2.39 & 0.47 & 6.95 & 0.20 & 23.54 \\
			
			9 & 2.58 & 0.50 & 7.43 & 0.21 & 24.51 \\
			
			10 & 2.77 & 0.53 & 7.83 & 0.22 & 25.37 \\
			
			11 & 2.97 & 0.56 & 8.13 & 0.24 & 26.11 \\
			
			12 & 3.16 & 0.58 & 8.41 & 0.25 & 26.62 \\
			
			13 & 3.35 & 0.61 & 8.72 & 0.25 & 26.97 \\
			
			14 & 3.54 & 0.63 & 8.98 & 0.27 & 27.47 \\
			
			15 & 3.73 & 0.65 & 9.15 & 0.27 & 27.69 \\
			
			16 & 3.92 & 0.67 & 9.32 & 0.28 & 27.83 \\
			
			17 & 4.11 & 0.69 & 9.53 & 0.29 & 28.08 \\
			
			18 & 4.31 & 0.70 & 9.66 & 0.30 & 27.99 \\
			
			19 & 4.50 & 0.72 & 9.86 & 0.31 & 28.04 \\
			
			20 & 4.69 & 0.73 & 9.98 & 0.31 & 28.02 \\
			
			21 & 4.88 & 0.74 & 10.04 & 0.32 & 27.89 \\
			
			22 & 5.07 & 0.75 & 10.14 & 0.33 & 27.78 \\
			
			23 & 5.26 & 0.76 & 10.25 & 0.33 & 27.62 \\
			
			24 & 5.45 & 0.77 & 10.36 & 0.34 & 27.49 \\
			
			25 & 5.64 & 0.78 & 10.45 & 0.34 & 27.30 \\
			
			26 & 5.83 & 0.79 & 10.49 & 0.35 & 26.96 \\
			
			27 & 6.03 & 0.80 & 10.53 & 0.35 & 26.76 \\
			
			28 & 6.22 & 0.80 & 10.57 & 0.36 & 26.55 \\
			
			29 & 6.41 & 0.81 & 10.55 & 0.36 & 26.24 \\
			
			30 & 6.60 & 0.81 & 10.51 & 0.37 & 25.98 \\
			
			31 & 6.79 & 0.82 & 10.45 & 0.37 & 25.73 \\
			
			32 & 6.98 & 0.82 & 10.43 & 0.37 & 25.39 \\
			
			33 & 7.17 & 0.83 & 10.40 & 0.38 & 25.09 \\
			
			34 & 7.36 & 0.83 & 10.41 & 0.38 & 24.82 \\
   
			35 & 7.55 & 0.83 & 10.35 & 0.38 & 24.45 \\      
			\hline
                \hline
		\end{tabular}
	\end{center}
\caption{J-factors values in the ROI. The columns show the
definition of the ith ROI together with the corresponding solid angle size, and values of the p- and d-wave J-factors obtained for DM-only and FIRE-2 simulations, respectively.}
 \label{tab:Jfactab}
\end{table*}
}

\clearpage

\section{Expected background fluxes in the GC region}
\label{sec:appB}
The energy-differential fluxes for the $\pi_0$ and ICS components for the baseline model adopted for the GDE predictions were introduced in Sec.~\ref{sec:GDEmodelling}. We show here the fluxes for the other three setups which are used to estimate the impact of the GDE modelling. 
The energy-differential fluxes for the $\pi_0$ and ICS components for all the other models considered for the GDE -- SA0 (R12), SA100 (F98) and SA0 (F98) -- are shown in Fig.~\ref{fig:fluxes_allgalprop}. As previously explained, we took the overall normalizations provided by the GALPROP predictions for SA100 (R12) and matched them to the measured values from \textit{Fermi}-LAT. The normalizations of the other three setups are rescaled such that the ratio of their normalizations to the baseline model one is maintained. We show the total flux in the GC region from our baseline model of the Galactic diffuse emission in Fig.~\ref{fig:hadronicMap}. This is obtained as a sum of the $\pi_0$ and ICS components at an energy of 10 TeV.

\begin{figure}[!ht] 
\includegraphics[width=0.45\textwidth]{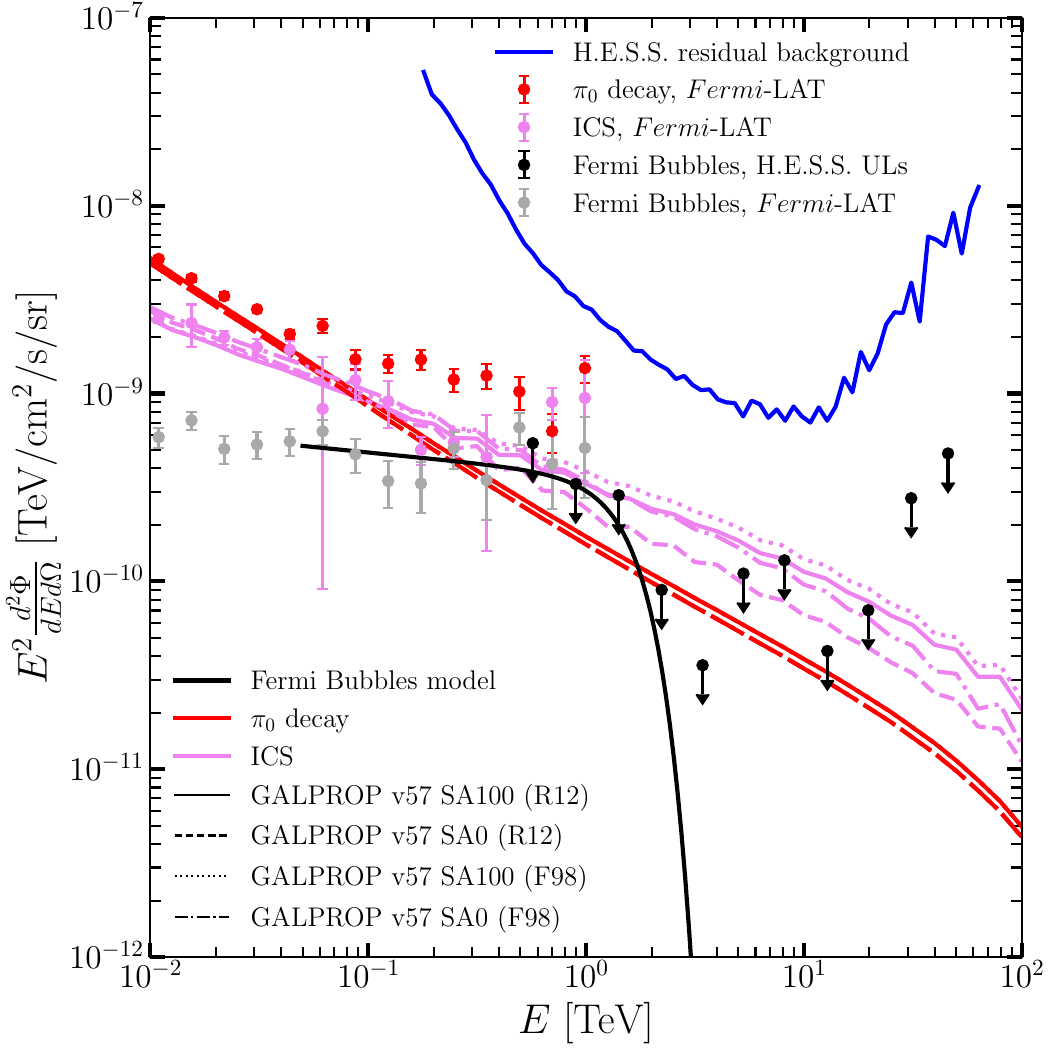}
\caption{Energy-dependent fluxes in  $E^2d^2\Phi/dEd\Omega$ versus energy $E$ for the backgrounds in the GC region.
The $\pi_0$ (pink line) and ICS (red line) components of the Galactic diffuse emission are obtained from the latest v57 GALPROP simulations for the ROI and all the models considered in this analysis -- SA100 (R12), SA0 (R12), SA100 (F98), and SA0 (F98) -- as solid, dashed, dotted and dashed-dotted lines, respectively. The red and violet points show the $\pi_0$ and ICS components, respectively, from \textit{Fermi}-LAT measurements.~\cite{Fermi-LAT:2017opo}. 
The residual background flux measured by H.E.S.S. is extracted from Ref.~\cite{HESS:2022ygk} and shown as a solid blue line. 
The flux points and upper limits from the \textit{Fermi}-LAT and H.E.S.S. analyses of the FBs, as described in Ref.~\cite{2021arXiv210810028M}, are shown as grey and black points, respectively. The best-fit spectrum of the FBs emission for energies above 100 GeV is shown as the solid black line~\cite{2021arXiv210810028M}.}
\label{fig:fluxes_allgalprop}
\end{figure}

\begin{figure}[!ht]
    \centering
    \includegraphics[scale=0.25]{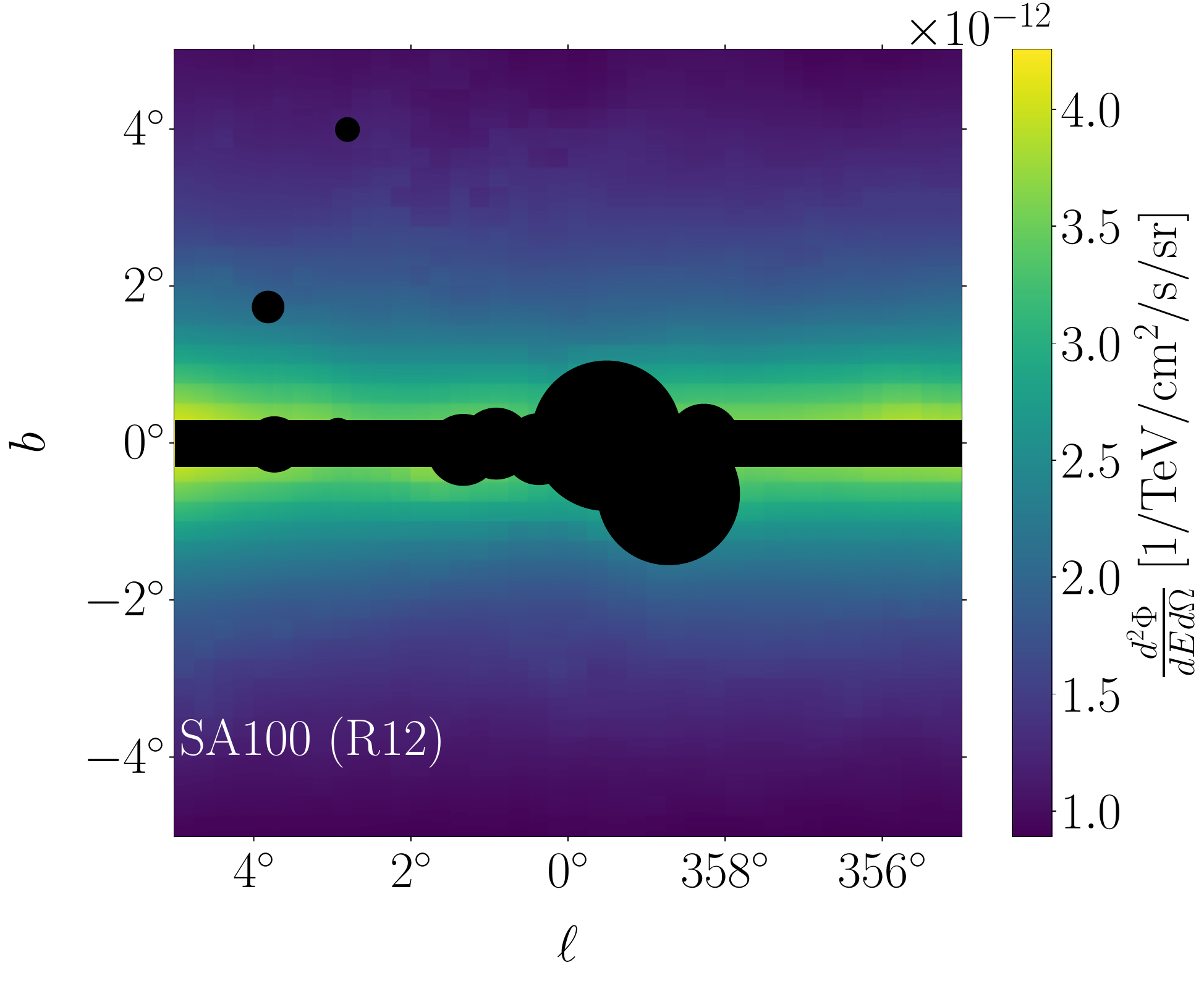}
    \caption{Total ($\pi_0$ decay + ICS) energy-differential gamma-ray flux map predicted by GALPROP v57~\cite{Porter:2021tlr} for a gamma-ray energy of 10 TeV displayed in Galactic coordinates ($l$,$b$). The black-filled areas correspond to the regions masked in the analysis. The simulation was performed with the new 3D interstellar gas model~\cite{Johannesson:2018bit} and the SA100 (R12) CR source density model. } 
    \label{fig:hadronicMap}
\end{figure}

\section{Mean expected upper limits and containment bands for p and d-wave annihilations}
\label{sec:appC}
The mean expected limits and confidence intervals are obtained by applying the Asimov procedure of Ref.~\cite{Cowan:2011an}. For the confidence intervals, to avoid limits below the expected one-sigma lower limit, we power constrain~\cite{Cowan:2011an}. Fig.~\ref{fig:LimitsContBands} show the power-constrained mean expected upper limits at 95\% C.L. on \sigmav~for p- and d-wave annihilations for the $W^+W^-$ channel, both when utilizing FIRE-2 and DM-only simulations for the DM distribution. Mean expected upper limits are obtained here without subtraction of the GDE contribution from the measured residual background. Containment bands at 1 and 2$\sigma$ are shown.

\begin{figure*}[!ht]
\includegraphics[width=0.44\textwidth]{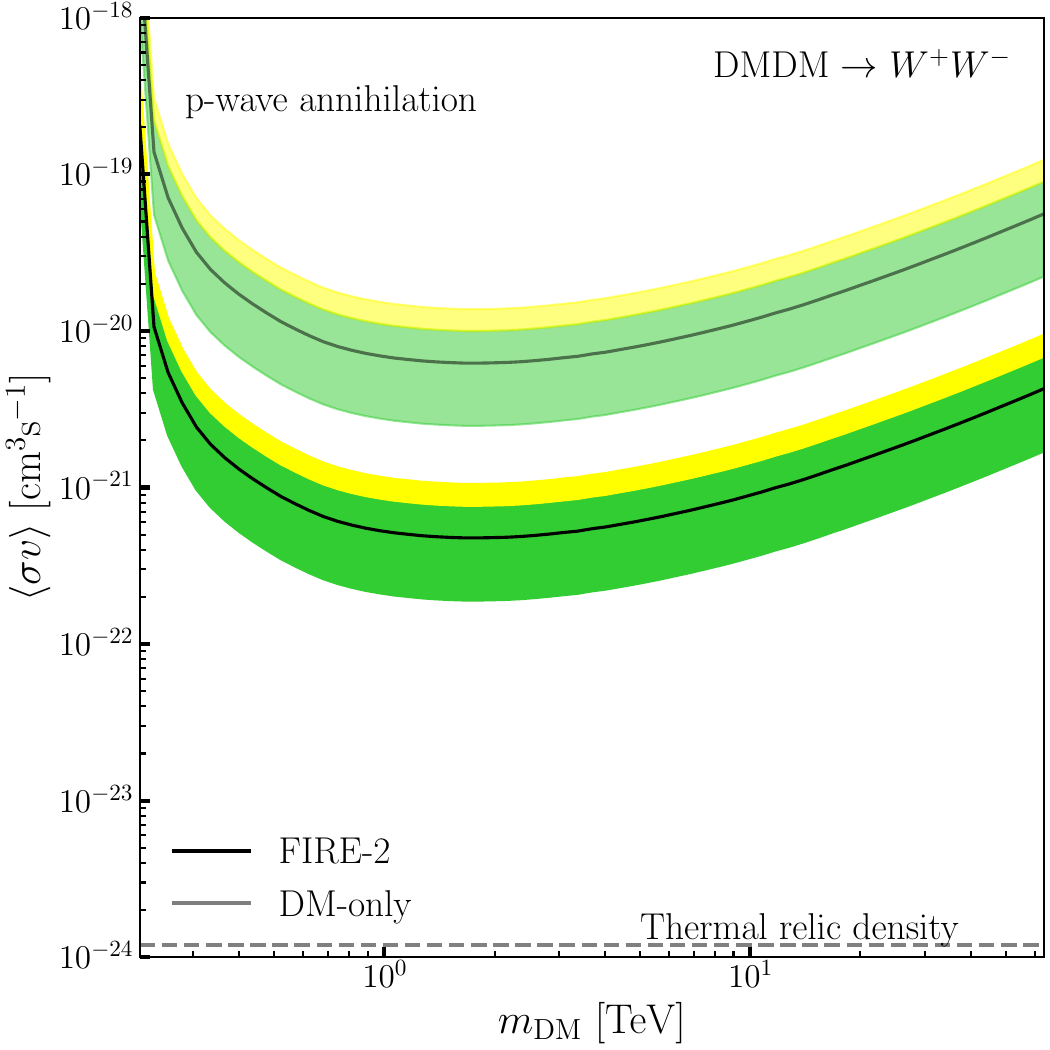}
\includegraphics[width=0.44\textwidth]{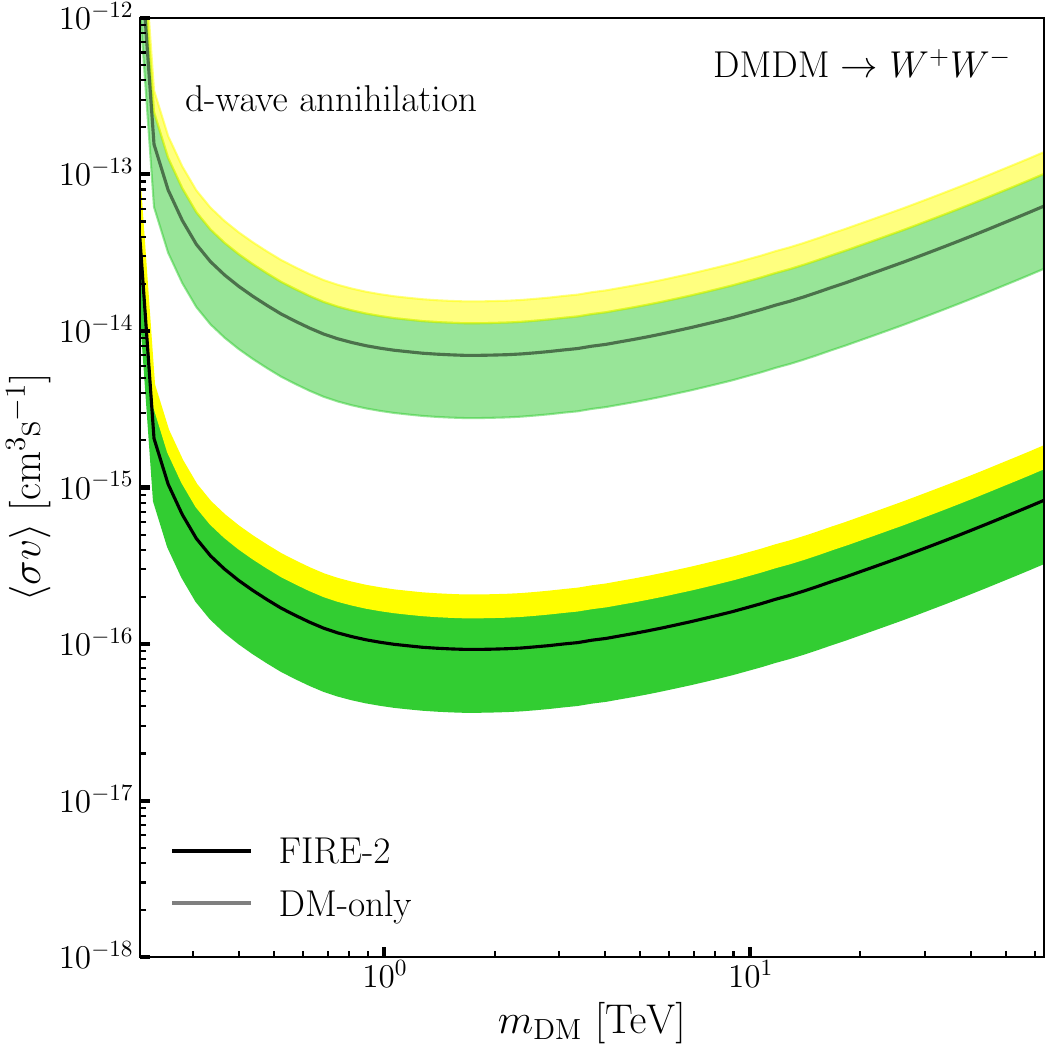}
\caption{Power-constrained mean expected upper limits computed at 95\% C.L. on \sigmav~for p-wave (letf panel) and d-wave (right panel) annihilation, respectively, for the $W^+W^-$ channel. The mean expected limits (solid black lines) are shown together with the 1 (green-shaded area) and 2$\sigma$ (yellow-shaded area) containment bands versus the DM mass for FIRE-2 (dark-shaded area) and DM-only (light-shaded area) simulations, respectively. The horizontal grey dashed line corresponds to the expected thermal annihilation cross section for the p-wave annihilation signal.}
\label{fig:LimitsContBands}
\end{figure*}

\clearpage

\bibliography{reference}	

\end{document}